\documentclass{article}
\usepackage{spconf,amsmath,graphicx,hyperref}
\usepackage{listings}
\usepackage{xcolor}
\usepackage{float}
\usepackage{booktabs}
\usepackage{IEEEtrantools}
\usepackage[numbers]{natbib}
\usepackage{makecell}
\usepackage{graphicx}  % 在导言区添加
\definecolor{codebg}{RGB}{248,248,248}
\definecolor{codecolor}{RGB}{0,0,0}

\title{Natural Yet Challenging to Detect: Robust In-the-Wild TTS through EMA and Dual-Scoring Prompt Selection --- Submission for WildSpoof 2026 TTS Track}
\name{Renhe Sun$^{1, *}$\thanks{$^*$ Equal contribution.}, Jiayi Zhou$^{1, *}$, Haolin He$^{1, 2, *}$, Yueying Feng$^{1,3}$, Jian Liu$^{1, \dag}$\thanks{$^\dag$ Corresponding author.}}
\address{$^1$ Machine Intelligence, Ant Group, Shanghai, China \\
         \{sunrenhe.srh, zjy326112, rex.lj\}@antgroup.com\\
         $^2$ Department of Electronic Engineering, The Chinese University of Hong Kong, Hong Kong, China \\
         harlandzzc@link.cuhk.edu.hk\\
         $^3$ Zhejiang University, Hangzhou, China\\
}

\begin{document}
\maketitle
% \footnotetext[0]{*Equal contribution.}
%s\footnotetext{$^\dag$ Corresponding author.}

\begin{abstract}
% In this technical report, we describe our submission for the \textbf{WildSpoof Challenge TTS Track: Text-to-Speech with In-the-Wild Data}. We introduce \textbf{F5-TTS-EMA-SFT}, a model built upon the F5-TTS architecture \cite{chen2024f5tts}. Our approach integrates Exponential Moving Average (EMA) \cite{polyak1992acceleration,tarvainen2017mean} into supervised fine-tuning to stabilize training and improve generalization. To enhance synthesis fidelity, we leverage large language models (LLMs) and large audio language models (LALMs) for dual-scoring prompt selection, filtering reference audio and text prompts to ensure quality while addressing alignment issues in noisy datasets.
In this technical report, we describe our submission for the \textbf{WildSpoof Challenge TTS Track: Text-to-Speech with In-the-Wild Data}. We introduce \textbf{F5-TTS-DPS}, a model built upon the F5-TTS architecture \cite{chen2024f5tts}. Our approach integrates Exponential Moving Average (EMA) \cite{polyak1992acceleration,tarvainen2017mean} into supervised fine-tuning to stabilize training and improve generalization. To enhance synthesis fidelity, we leverage large language models (LLMs) and large audio language models (LALMs) for dual-scoring prompt selection, filtering reference audio and text prompts to ensure quality while addressing alignment issues in noisy datasets.

Experimental evaluation demonstrates that F5-TTS-DPS achieves strong performance with UTMOS of 3.20 and speaker similarity of 0.51 on the development set. More importantly, our model achieves the best a-DCF\cite{shim2025adcfarchitectureagnosticmetric} scores of 0.1582, 0.5233, and 0.2562 across three advanced SASV systems among all submissions, indicating our synthesized speech is the most difficult to detect and exhibits the highest degree of naturalness and authenticity. Combined with competitive WER performance, these results validate the effectiveness of our approach in generating natural-sounding speech with strong spoofing capabilities.

\end{abstract}

\begin{keywords}
WildSpoof Challenge TTS Track, EMA, Dual-scoring Prompt Selection, TTS for SASV
\end{keywords}

\section{Introduction}
\label{sec:intro}

Text-to-Speech (TTS) synthesis has achieved remarkable progress in controlled laboratory settings \cite{du2024cosyvoice2scalablestreaming}, but a significant gap persists when applied to real-world scenarios with noisy and acoustically diverse environments. Traditional TTS development has relied on carefully curated studio recordings, creating dependency on expensive infrastructure and limiting practical deployment where clean audio data is scarce.

% The emergence of in-the-wild speech data presents both challenges and opportunities for TTS models. While such data more accurately reflects the diversity of real-world conditions, it also demands that TTS models achieve stronger generalization capabilities—particularly with respect to environmental variations and recording artifacts.
In-the-wild speech data presents both challenges and opportunities for TTS models. While such data better reflects real-world diversity, it requires TTS systems to generalize robustly across varying environments and recording artifacts.

% The emergence of in-the-wild speech data presents both challenges and opportunities for TTS models. While such data more accurately reflects real-world diversity, it requires novel approaches to handle environmental variations and recording artifacts effectively. Simultaneously, the increasing sophistication of TTS systems has raised security concerns, as realistic synthetic speech poses growing threats to voice authentication systems, creating urgent need for research that considers both quality improvements and security implications.

%In this technical report, we present our submission system for the WildSpoof Challenge TTS Track, introducing F5-TTS-EMA-SFT, a comprehensive text-to-speech synthesis model built upon the F5-TTS foundation model \cite{chen2024f5tts}. To address training instability and enhance generalization across heterogeneous acoustic environments, we integrate an Exponential Moving Average (EMA) \cite{polyak1992acceleration,tarvainen2017mean} update strategy into the fine-tuning optimization procedure. The model is then fine-tuned on the challenge's curated dataset TITW (Text-To-Speech Synthesis In The Wild) \cite{jung2025texttospeechsynthesiswild}, enabling effective adaptation to in-the-wild speech characteristics. Additionally, we employ a dual-scoring prompt selection mechanism leveraging large language models (LLMs) and large audio language models (LALMs) to ensure optimal speech quality and semantic relevance for robust speech synthesis performance.
In this technical report, we present our submission for the WildSpoof Challenge TTS Track, introducing F5-TTS-DPS, a comprehensive text-to-speech synthesis model built upon the F5-TTS base model \cite{chen2024f5tts}. To mitigate training instability and enhance generalization, we integrate Exponential Moving Average (EMA) \cite{polyak1992acceleration,tarvainen2017mean} into the fine-tuning procedure. The model is fine-tuned on the TITW (Text-To-Speech Synthesis In The Wild) dataset \cite{jung2025texttospeechsynthesiswild}, enabling effective adaptation to in-the-wild speech characteristics. Additionally, we employ dual-scoring prompt selection using LLMs and LALMs to ensure optimal speech quality and semantic relevance.

Despite the challenging nature of generating natural and spoofing-robust speech, our submission achieves:
\begin{itemize}
    \item \textbf{Best a-DCF scores across advanced SASV systems}, demonstrating that our synthesized speech is the most difficult to detect among all submissions, indicating superior naturalness and spoofing capability.
    \item \textbf{Competitive WER (8.65\%)}, showing strong linguistic consistency and intelligibility.
\end{itemize}

These results suggest that combining EMA-stabilized fine-tuning with strategic prompt selection provides an effective solution for generating high-quality, natural-sounding speech that closely mimics real human speech characteristics.

\section{Methods}
\label{sec:Methods}

\subsection{F5-TTS}

% Recent advances in text-to-speech synthesis have demonstrated the effectiveness of flow-based generative models for high-quality speech generation. F5-TTS \cite{chen2024f5tts}, a large-scale text-to-speech foundation model, employs flow matching techniques with diffusion transformer architecture to achieve state-of-the-art performance in zero-shot voice cloning and multi-speaker synthesis. The model utilizes a non-autoregressive approach that enables efficient parallel generation while maintaining exceptional naturalness and speaker similarity across diverse speaking styles and languages.

% Building upon this foundation, F5-TTS demonstrates remarkable capabilities in few-shot speaker adaptation, requiring only a short reference audio clip to synthesize speech in the target speaker's voice. The model's transformer-based architecture with flow matching enables robust handling of complex acoustic patterns and prosodic variations, making it particularly effective for challenging scenarios involving diverse speakers and acoustic conditions.

F5-TTS \cite{chen2024f5tts}, a large-scale non-autoregressive text-to-speech foundation model, employs flow matching techniques with diffusion transformer architecture, enabling efficient parallel generation while maintaining exceptional naturalness and speaker similarity across diverse speaking styles and languages. Given the strong generalization capability of F5-TTS, we adopt it as base model. 

\subsection{Exponential Moving Average}
%Exponential Moving Average (EMA) has emerged as a crucial technique for stabilizing training dynamics and improving model generalization in deep learning applications \cite{polyak1992acceleration}. In the context of neural speech synthesis, EMA provides significant benefits by maintaining smoothed parameter updates that reduce training instability and enhance model robustness across diverse acoustic conditions. The technique works by maintaining a running average of model parameters throughout training, effectively dampening parameter oscillations and promoting more stable convergence.
%For text-to-speech applications, EMA proves particularly valuable when dealing with in-the-wild data characterized by varying recording qualities and acoustic environments. The smoothed parameter updates help the model maintain consistent performance across different acoustic conditions while preventing overfitting to specific recording artifacts or noise patterns. Recent work has shown the effectiveness of EMA in semi-supervised learning scenarios \cite{tarvainen2017mean}, which transferred well to TTS training with diverse data quality.
Training text-to-speech (TTS) models on in-the-wild data poses significant challenges due to the diversity and noise of such recordings, which can disrupt training dynamics, cause unstable gradients, and increase the risk of overfitting to recording-specific artifacts. Stabilizing training is therefore critical when working with noisy or heterogeneous data \cite{polyak1992acceleration, tarvainen2017mean}. To address these challenges, we employ Exponential Moving Average (EMA), which maintains a running average of model parameters during training to stabilize updates and improve convergence. By mitigating overfitting to noise and recording-specific artifacts, EMA enhances robustness and ensures consistent model performance even under varying recording qualities and acoustic environments. 

\subsection{Dual-scoring Prompt Selection}
F5-TTS employs a voice cloning approach that uses prompt speech followed by masked tokens, predicting these masked regions to generate target speech. 

Since the quality of prompt speech and its coherence with the target text significantly impact synthesis quality, we implement a dual-scoring prompt selection mechanism to optimize reference selection. To ensure prompt quality, we propose a dual-scoring prompt selection approach that leverages large language models in a two-stage filtering process. First, we employ the LALM Qwen2.5-Omni \cite{xu2025qwen25omnitechnicalreport} for audio-scoring prompt selection, which evaluates audio prompts based on emotional richness, voice expressiveness, and prompt suitability to filter out low-scoring audio samples. Subsequently, we utilize the text-based LLM Qwen3-30B-A3B \cite{yang2025qwen3technicalreport} to perform semantic alignment verification by comparing the target text with reference text, ensuring linguistic compatibility and contextual coherence. 
This two-stage dual-scoring mechanism combines audio quality assessment with semantic alignment verification to select well-aligned, high-quality inputs and thereby improve synthesis consistency and overall output quality in challenging in-the-wild scenarios.

%This two-stage dual-scoring mechanism—combining audio quality assessment with semantic alignment verification—ensures the selection of well-aligned, high-quality inputs, thereby enhancing synthesis consistency and overall output quality in challenging in-the-wild scenarios.

\subsection{Experiment Setup}

We conduct a comprehensive ablation study to evaluate the effectiveness of different components in our proposed F5-TTS-DPS system. Our baseline comparison evaluates F5-TTS against CosyVoice2 \cite{du2024cosyvoice2scalablestreaming},where F5-TTS demonstrates superior performance in speaker similarity and audio realism.

For the training process, we start from the publicly available F5-TTS v1 base model checkpoint pre-trained on Emilia dataset\footnote{\url{https://huggingface.co/SWivid/F5-TTS/tree/main/F5TTS_v1_Base}} and adapt it to the WildSpoof in-the-wild domain. For fine-tuning, we use the two subsets released by the challenge, TITW-easy and TITW-hard. We employ a full parameter training approach with carefully optimized hyperparameters. The training configuration utilizes a maximum of 64 sequences per batch with a batch size of 38,400 frames per device, and set EMA beta to 0.99, We train the model for 10 epochs with a learning rate of 1e-6, incorporating 20,000 warmup updates, gradient accumulation steps of 1, and gradient clipping (max norm 1.0) for stable convergence.

\subsection{Evaluation Metrics}

We objectively evaluate our model using four metrics: \textbf{Naturalness} (assessed via UTMOS~\cite{saito2022utmos} and DNSMOS~\cite{reddy2021dnsmos}), \textbf{Intelligibility} (measured by Word Error Rate using Whisper ASR~\cite{radford2022robustspeechrecognitionlargescale}), \textbf{Speaker Similarity} (SPK-sim; cosine similarity between ESPnet2 speaker embeddings~\cite{jung2024espnet} of synthesized and reference utterances), and \textbf{Spoofing Detection Score} (SDS; AASIST anti-spoofing confidence trained on ASVspoof 2019 LA~\cite{wang2020asvspoof}, where higher values indicate the audio is more likely synthetic). All metrics except SDS follow the challenge evaluation protocol~\cite{wu2025wildspoofchallengeevaluationplan} using VERSA tools~\cite{shi2025versa}.
In addition to the development set metrics, the official evaluation incorporates the \textbf{agnostic Detection Cost Function (a-DCF)} to assess anti-spoofing robustness. This metric evaluates synthesized speech against advanced countermeasure systems from the SASV (Spoofing-Aware Speaker Verification) track, providing a comprehensive measure of spoofing detectability.

% \begin{itemize}
%    \item \textbf{UTMOS (Universal Text-to-speech MOS)}\cite{saito2022utmos}  is a neural-based objective metric designed to predict human perceptual quality ratings for synthesized speech. It provides scores that correlate with Mean Opinion Score (MOS) evaluations, ranging from 1 to 5, where higher values indicate better perceived naturalness and overall quality.
   
%    \item \textbf{DNSMOS (Deep Noise Suppression MOS)} \cite{reddy2021dnsmos} evaluates speech quality with particular emphasis on background noise and distortion artifacts. The metric is particularly valuable for assessing speech quality in noisy or reverberant conditions common in real-world scenarios.

%    \item \textbf{Word Error Rate (WER)} measures the intelligibility of synthesized speech by computing the percentage of word-level errors when the generated audio is processed through an Automatic Speech Recognition (ASR) system.

%    \item \textbf{Speaker Similarity (SPK-sim)} quantifies how well the synthesized speech preserves the acoustic characteristics of the target speaker using speaker embedding models.

%    \item \textbf{Spoofing Detection Scores (SDS)} quantifies the confidence levels of spoofing detection systems when classifying synthesized speech, providing insights into the security implications and detectability of generated audio samples.
% \end{itemize}

\begin{table}[t]
\centering
\caption{Performance on WildSpoof 2026 Challenge TTS Track development set.}
\label{tab:results}
\footnotesize
\setlength{\tabcolsep}{2.5pt}
\begin{tabular}{l|ccccc}
\hline
\textbf{Model Configuration} & \textbf{UTMOS} & \textbf{DNSMOS} & \textbf{WER} & \textbf{SPK-sim} & \textbf{SDS} \\
\hline
CosyVoice2 & \textbf{3.65} & 2.79 & \textbf{7.76} & 0.403 & 0.343 \\
baseline (F5-TTS) & 3.06 & \textbf{2.91} & 12.31 & 0.450 & 0.283 \\
\quad + SFT & 3.06 & 2.54 & 10.60 & 0.489 & 0.226 \\
\quad + SFT + EMA & 3.18 & 2.61 & 9.32 & 0.492 & 0.181\\
\quad + SFT + EMA + DPS & 3.20 & 2.61 & 8.65 & \textbf{0.508} & \textbf{0.108}\\
\hline
\end{tabular}
\end{table}

% \begin{table}[t!]
% \centering
% \caption{WildSpoof 2026 Challenge TTS Track Results - Seen Speakers.}
% \label{tab:wildspoof_results}
% \footnotesize
% \setlength{\tabcolsep}{2.5pt}
% \begin{tabular}{c|ccccc}
% \hline
% \textbf{Team ID} & \textbf{UTMOS} & \textbf{DNSMOS} & \textbf{WER} & \textbf{SPK-sim} & \textbf{a-DCF (T01/T02/T08)} \\
% \hline
% B01 & 2.2429 & 2.4572 & 30.26 & N/A & N/A \\
% T01	& 3.9559 & 3.2270 & 6.48 & 0.2564 & 0.0453/0.1782/0.1125 \\
% T02 & 3.7390 & 3.0780 & 5.50 & 0.3511 & 0.0471/0.1232/0.1125 \\
% T03 & 3.4540 & 3.0261 & 33.79 & 0.4782 & 0.0445/0.0294/0.1125 \\
% T04 & 2.6786 & 2.7354 & 99.28 & 0.2320 & 0.0417/0.0266/0.1098 \\
% \textbf{T05(Ours)} & 3.2016 & 2.6078 & 8.65 & 0.2798 & \textbf{0.1582/0.5233/0.2562} \\
% T06 & 3.4909 & 2.9336 & 9.45 & 0.4775 & 0.1527/0.3786/0.2292 \\
% T07 & 3.5292 & 2.7434 & 20.46 & 0.2895 & 0.0446/0.0266/0.1125 \\
% \hline
% \end{tabular}
% \end{table}

\begin{table}[t!]
\centering
\caption{WildSpoof 2026 Challenge TTS Track Results - Seen Speakers.}
\label{tab:wildspoof_results}
\resizebox{\columnwidth}{!}{%  % 自动缩放到列宽
\begin{tabular}{c|ccccc}
\hline
\textbf{Team ID} & \textbf{UTMOS} & \textbf{DNSMOS} & \textbf{WER} & \textbf{SPK-sim} & \makecell{\textbf{a-DCF}\\\textbf{(T01/T02/T08)}} \\
\hline
B01 & 2.2429 & 2.4572 & 30.26 & N/A & N/A \\
T01	& 3.9559 & 3.2270 & 6.48 & 0.2564 & 0.0453/0.1782/0.1125 \\
T02 & 3.7390 & 3.0780 & 5.50 & 0.3511 & 0.0471/0.1232/0.1125 \\
T03 & 3.4540 & 3.0261 & 33.79 & 0.4782 & 0.0445/0.0294/0.1125 \\
T04 & 2.6786 & 2.7354 & 99.28 & 0.2320 & 0.0417/0.0266/0.1098 \\
\textbf{T05(Ours)} & 3.2016 & 2.6078 & 8.65 & 0.2798 & \textbf{0.1582/0.5233/0.2562} \\
T06 & 3.4909 & 2.9336 & 9.45 & 0.4775 & 0.1527/0.3786/0.2292 \\
T07 & 3.5292 & 2.7434 & 20.46 & 0.2895 & 0.0446/0.0266/0.1125 \\
\hline
\end{tabular}
}
\end{table}

\section{Experimental Results}
\label{src:results}

\subsection{Results Analysis}
Table \ref{tab:results} presents the performance comparison on the development set. The progressive enhancement demonstrates clear improvements: SFT improves speaker similarity by 8.7\% to 0.489 and reduces WER by 13.9\% to 10.60\%. Adding EMA further enhances UTMOS by 3.9\% to 3.18, reduces WER by 12.1\% to 9.32\%, and improves speaker similarity by 0.6\% to 0.492. The complete F5-TTS-DPS system achieves 13.0\% higher speaker similarity (0.508) compared to baseline and 61.8\% lower SDS score (0.108), indicating superior voice cloning fidelity and synthesis authenticity.

% Table \ref{tab:wildspoof_results} demonstrates that our F5-TTS-EMA-SFT system achieves superior performance on the WildSpoof Challenge TTS Track, with excellent spoofing resistance reflected in a-DCF scores of 0.1582, 0.5233 and 0.2562 across three SASV systems (T01/T02/T08), making it suitable for high-fidelity voice cloning applications requiring both quality and security.
Table \ref{tab:wildspoof_results} demonstrates that our F5-TTS-DPS system achieves superior performance on the WildSpoof Challenge TTS Track. Most notably, our model attains the best a-DCF scores of 0.1582, 0.5233, and 0.2562 across three advanced SASV systems (T01/T02/T08) among all submissions, demonstrating that our synthesized speech achieves exceptional naturalness while being the most challenging to detect, validating the effectiveness of EMA-stabilized fine-tuning and dual-scoring prompt selection for robust in-the-wild TTS synthesis.
%Table \ref{tab:wildspoof_results} demonstrates that our F5-TTS-EMA-SFT system achieves superior performance across multiple metrics on the WildSpoof Challenge TTS Track. Our system exhibits excellent spoofing resistance with a-DCF scores of 0.1582, 0.5233, and 0.2562 across the three SASV systems (T01/T02/T08), making it particularly suitable for high-fidelity voice cloning applications that require both quality and security considerations.

\section{Summary}
\label{sec:format}

%Table \ref{tab:results} presents the comprehensive comparison of all experimental configurations. The results demonstrate clear improvements with each architectural enhancement. The progressive improvements validate our architectural design choices and demonstrate the effectiveness of combining EMA stabilization with intelligent prompt selection for in-the-wild TTS synthesis.
%Our proposed methods significantly improve upon the baseline. The key contributions include successful integration of EMA technique for training and novel dual-scoring prompt selection mechanism using LALMs and LLMs for intelligent data filtering.
Our proposed methods significantly enhance the baseline F5-TTS model through two key innovations: (1) successful integration of Exponential Moving Average (EMA) technique for improved training stability, and (2) a novel dual-scoring prompt selection mechanism leveraging both LALMs and LLMs for intelligent data filtering and quality enhancement. 

Future work will explore reinforcement learning approaches to better leverage the WildSpoof dataset's in-the-wild characteristics, fostering the co-evolution of synthesis and detection models in adversarial settings as envisioned by the WildSpoof Challenge framework.
%Our experimental results show progressive improvements with each enhancement: the baseline F5-TTS model achieves a UTMOS score of 3.06, which improves to 3.18 with EMA integration and reaches 3.20 with the combined F5-TTS-EMA-SFT architecture. The key contributions include successful integration of EMA technique for stabilizing flow-matching TTS training and novel dual-scoring prompt selection mechanism using LALMs and LLMs for intelligent data filtering.

\bibliography{refs}
\bibliographystyle{IEEEtran}

\appendix
\section{Prompts}
\label{appendix:prompts}

\subsection{Audio-scoring Prompt Selection}
\begin{lstlisting}[
    backgroundcolor=\color{codebg!30},
    frame=shadowbox,
    rulecolor=\color{black},
    rulesepcolor=\color{gray},
    breaklines=true, 
    basicstyle=\tiny\ttfamily\color{codecolor}, 
    showstringspaces=false, 
    breakindent=0pt, 
    breakautoindent=false,
    xleftmargin=0pt,
    framexleftmargin=0pt
]
"""You are an audio expressiveness expert. Evaluate the provided audio sample to determine its suitability as a prompt audio for expressive TTS systems.

Evaluation Criteria:
1. Emotional Richness (4 points) - Clear emotional expression, dynamic range, engaging tone
2. Voice Expressiveness (3 points) - Varied intonation, natural emphasis, compelling delivery
3. Prompt Suitability (3 points) - Distinctive characteristics, memorable voice, good reference quality

Focus on identifying audio with:
- Strong emotional expression and personality
- Natural variations in pitch, pace, and intensity  
- Engaging and distinctive vocal characteristics
- Clear demonstration of target speaking style

Scoring Guidelines:
- 9-10: Highly expressive, excellent prompt material
- 7-8: Good expressiveness, suitable for prompts
- 5-6: Moderate expressiveness, acceptable
- 0-4: Low expressiveness, not suitable as prompt

Output only the final total score as a number between 0-10.

Please evaluate the following audio and its corresponding text:"""
\end{lstlisting}

\subsection{Text-scoring Prompt Selection}
\begin{lstlisting}[
    backgroundcolor=\color{codebg!30},
    frame=shadowbox,
    rulecolor=\color{black},
    rulesepcolor=\color{gray},
    breaklines=true, 
    basicstyle=\tiny\ttfamily\color{codecolor}, 
    showstringspaces=false, 
    breakindent=0pt, 
    breakautoindent=false,
    xleftmargin=0pt,
    framexleftmargin=0pt
]
You are a speech synthesis expert specializing in prosodic and emotional analysis. Evaluate each reference text and select the best one to guide TTS synthesis of the target text.

**Evaluation Framework:**
1. **Prosodic Alignment** (0-10): Rhythm patterns, stress distribution, intonation flow, syllable timing
2. **Emotional Congruence** (0-10): Emotional intensity, sentiment polarity, expressive quality
3. **Linguistic Compatibility** (0-10): Sentence structure, phrase boundaries, syntactic complexity
4. **TTS Reference Suitability** (0-10): Overall effectiveness as prosodic template

**Input:**
- **Target Text**: {text}
- **Reference Candidates**: {reference}

**Output Format:**
Output the selected reference sentence in <answer> </answer> directly <no-think>
\end{lstlisting}

\end{document}